\documentstyle[12pt,epsfig]{article} 
\newlength{\dinwidth}                       
\newlength{\dinmargin}                      
\newcommand{\pmb}[1]{%
        \setbox0=\hbox{#1}%
        \kern-.02em\copy0\kern-\wd0
        \kern+.04em\copy0\kern-\wd0
        \kern-.02em\raise.0217em\box0}

\newcommand{\lsim}{
 \mathrel{\setbox0=\hbox{$<$}\raise0.6ex\copy0\kern-\wd0
 \lower0.65ex\hbox{$\sim$}}}
\newcommand{\gsim}{
 \mathrel{\setbox0=\hbox{$>$}\raise0.6ex\copy0\kern-\wd0
 \lower0.65ex\hbox{$\sim$}}}

\begin{document}

\centerline{\Large\bf  Nearthreshold Large $Q^2$ Electroproduction}
\centerline{\Large\bf  off Polarized Deuteron}

\vspace{0.4cm}
\centerline{\large L.~Frankfurt$^{a,d}$, 
M.~Sargsian$^{a,e}$, M.~Strikman$^{b,d}$}

\vspace{-0.4cm}

\begin{quotation}
\noindent $^{(a)}$ Tel Aviv University, Tel Aviv, Israel \\
\noindent $^{(b)}$ Pennsylvania State University, University Park, PA, USA \\
\noindent $^{(d)}$ S.Petersburg Nuclear Physics Institute, Russia \\ 
\noindent $^{(e)}$ Yerevan Physics Institute, Yerevan, Armenia  
\end{quotation}

\vspace{-0.6cm}

\begin{quotation}
\noindent
{\bf Abstract:}
The exclusive and inclusive electroproduction off the polarized deuteron  
is considered at large $Q^2$ and $x \ge 0.5$.
It is shown that the use of a polarized target will allow to emphasize 
smaller than average internucleon distances in the deuteron. 
As a result, we expect amplification of all the effects (color transparency, 
relativistic dynamics, etc.) sensitive to small internucleon distances. 
Numerical estimates are given for the processes $e+ \vec d \rightarrow e +p+n$
and $e+ \vec d \rightarrow e +X$.
\end{quotation}

\vspace{-0.2cm}

\noindent{\Large\bf 1 \  Motivation}

The theoretical analysis \cite{FGMSS95} of the intermediate energy 
$Q^2\sim 1$ GeV$^2$ electrodisintegration of the deuteron at
$x\sim 1$ indicates that there is a fast convergence of the 
higher (large $l$) partial waves of the final $pn$ continuum wave function. 
As a result, we can substitute the (infinite) sum over the 
partial waves with the phenomenological amplitude for $pn$ scattering.
This simplification allows to implement relativistic kinematics
of the final state interaction (FSI) amplitude through the analysis 
of the corresponding (covariant) Feynman diagrams \cite{FSS96}.
The main theoretical conclusion \cite{FSS96} is that, at $Q^2\ge 1$ GeV$^2$,
there exists a unique scheme of legitimate calculations within the extended 
eikonal approximation which selfconsistently accounts for relativistic 
dynamics.  This enhances considerably the exploration  potential of the 
electroproduction reaction, especially off a deuteron target, whose 
wave function is well established at Fermi momenta $\le 400$ MeV/c.  

\noindent
Based on this, we discuss two alternative studies: 

$\bullet$ Investigation of the QCD prediction that the absorption of a
high momentum virtual photon by a nucleon leads to the production of a
small size color singlet state, optimistically called a point-like 
configuration (PLC). Such a study requires selection of kinematics where 
small enough Fermi momenta dominate and where the transverse momenta of 
the spectator nucleons are large enough so that the dominant contribution 
is given by the reinteraction of the PLC with a spectator nucleon 
(see Sect. 2).

$\bullet$  Probing relativistic effects in deuteron electrodisintegration at  
moderate $Q^2\le 4$ GeV$^2$ and rather large longitudinal Fermi momenta.  
Such a study will provide a critical discrimination between the different 
approaches to high energy scattering off deeply bound nucleons.

Both these  studies  would greatly benefit from the use of a polarized 
target.  The reason is that the use of a $\vec d$ allows to enhance 
the contribution of the $D$-state in the deuteron's ground state wave 
function.  Due to the diminishing probability of the $D$-state at small Fermi 
momenta, these reactions would be sensitive to smaller internucleon distances 
in the deuteron as compared to the unpolarized case, leading to an
amplification of all the effects sensitive to small internucleon distances.

\medskip

\noindent{\Large\bf 2 \  Color Transparency Effects and Vanishing FSI}

In QCD, the absorption of a high $Q^2$ photon by a nucleon produces a PLC,
which, at very high energies, would not interact with the nucleons, thus 
eliminating FSI.  This vanishing of
\newpage

\begin{figure}[t]
\vspace{1.8cm}
\centering{
\epsfig{figure=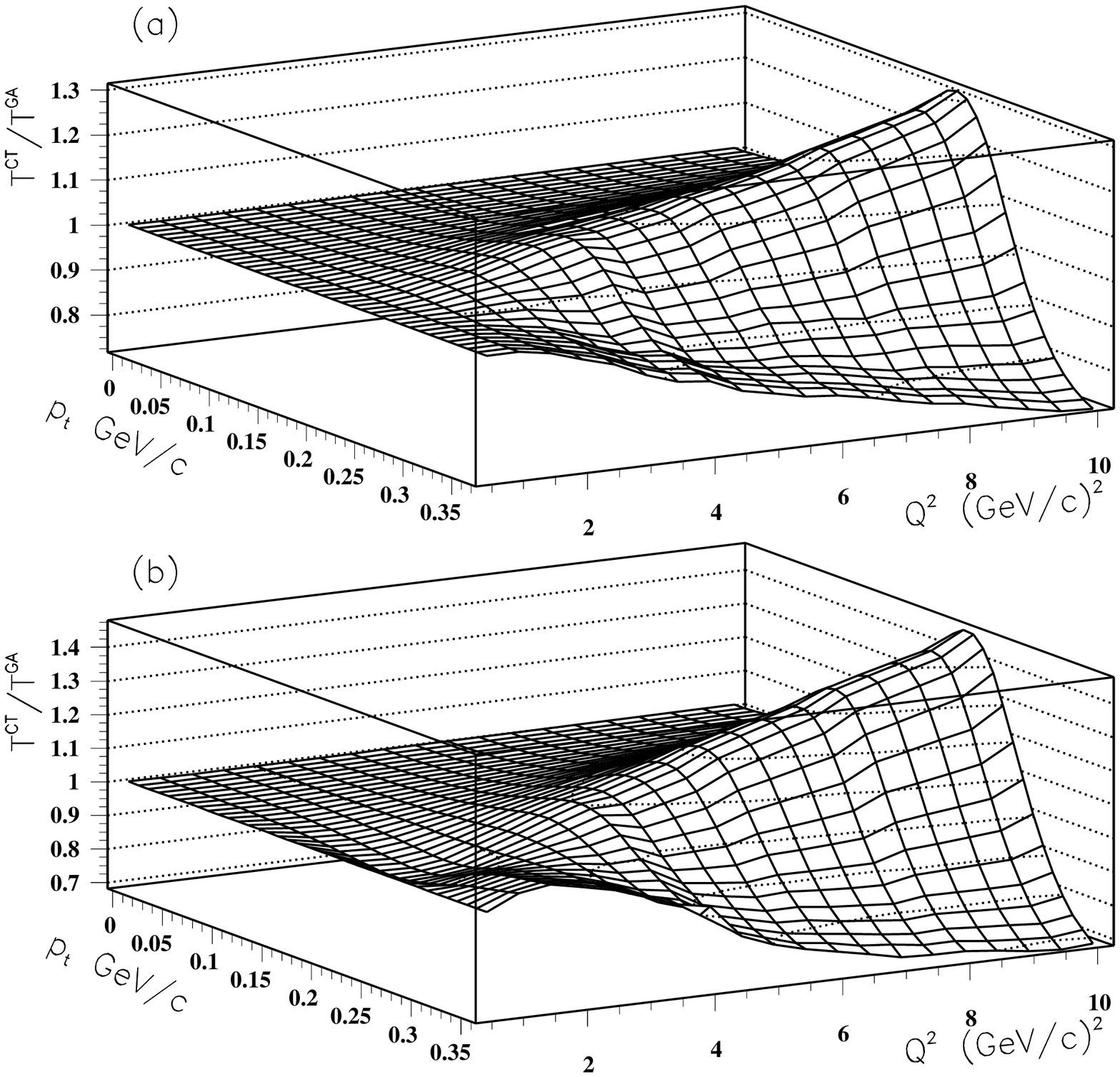,height=8.6cm}}
\end{figure}
\begin{figure}[t]
\vspace{-8.6cm}
\hspace{8.4cm}
\centering{
\epsfig{figure=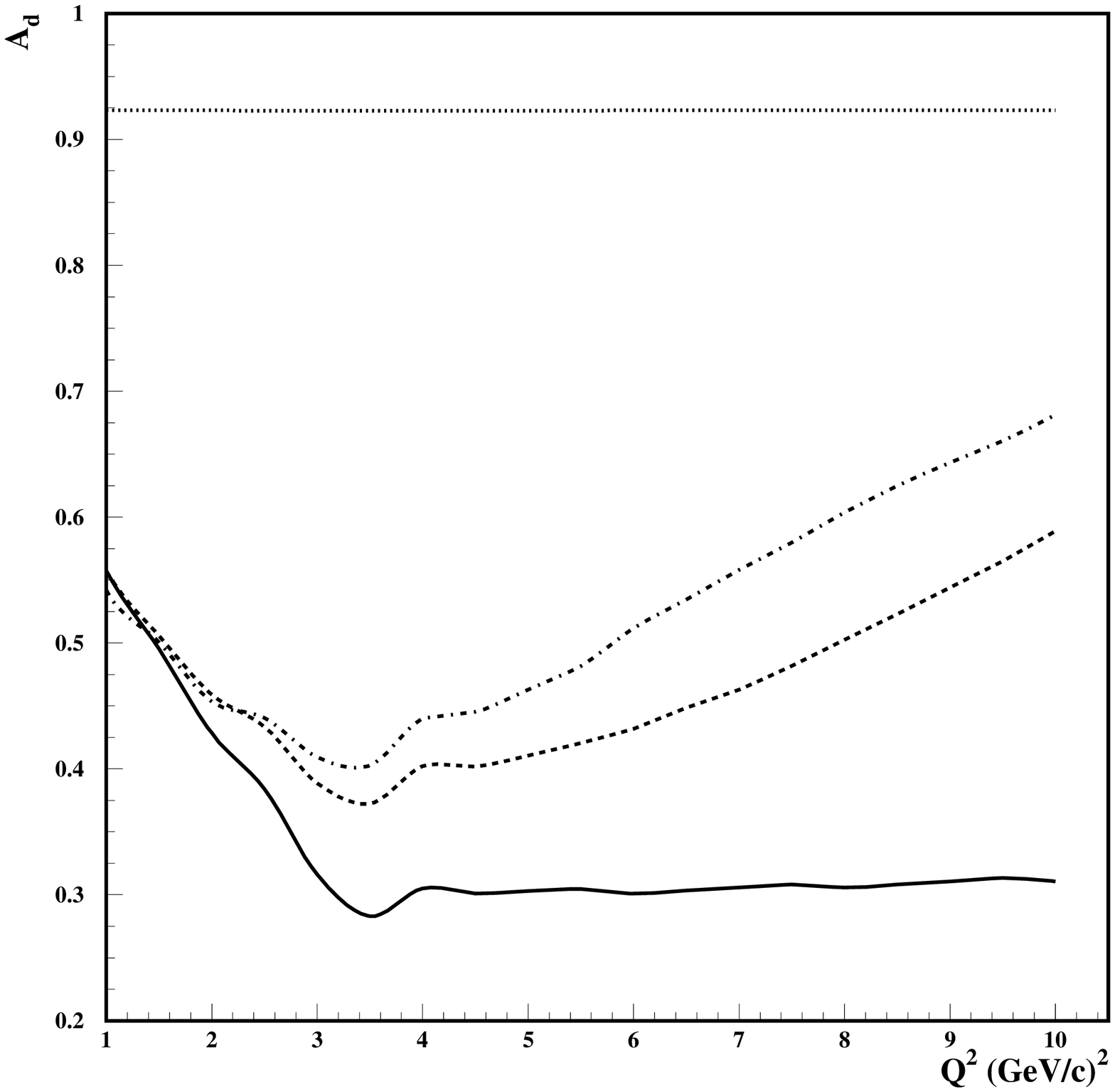,height=8.0cm}}
\end{figure}

\begin{picture}(100,200)(0,0)
\put(-10,255){{\bf Figure 1:} {\em $p_t$ and $Q^2$ dependence of the ratio}} 
\put(-10,240){{\em $T^{GA}/T^{CT}$ for $\alpha \equiv (E_s-p_s^z)/m = 1$.}}
\put(-10,225){{\em a) quantum diffusion, b) three state model.}} 
\put(240,255){{\bf Figure 2:} {\em $Q^2$ dependence of $A_d$ for $\alpha =1$.}} 
\put(240,240){{\em Solid line - elastic eikonal, dashed - QDM,}} 
\put(240,225){{\em dashed-dotted - three state model, dotted -}}
\put(240,210){{\em PWIA.}}
\end{picture}

\vspace{-7.2cm}

\noindent 
the FSI has been termed color transparency (CT). 
At high but finite energies, a PLC is actually produced, but it  
expands as it propagates through the nucleus \cite{ANN}.
To suppress the expansion effects, it is necessary to ensure that 
the expansion length, $l_h\sim 0.4(p/$GeV), is greater than the
characteristic longitudinal distance in  the reaction.
In the considered $d(e,e'pn)$ and $d(e,e'pN^*)$ reactions, where one nucleon
carries almost all the momentum of the photon while the second nucleon 
(or its resonance) is a spectator, the actual expansion
distances are the distances between the nucleons in the
deuteron \cite{FGMSS95}. Thus, suppressing large distance effects
through the deuteron's polarization, one effectively will diminish the 
PLC's expansion, leading to an earlier onset of CT.

The scattering amplitude
${\cal M}$, including the $np$ final state interaction, can be written 
as: 
\vspace{-0.2cm}
\begin{equation}
{\cal M} = <p_s^z,\vec p_t|d> -{1\over 4i}\int{d^2k_t\over 
(2\pi)^2}<\tilde p_s^z,\vec p_t-\vec k_t|d>
 {\bf F^{np}(\vec k)}\left[1-i\beta\right],
\label{eq:amp}
\end{equation}
where $\tilde p_s^z=p_s^z-(E_s-m){M_d+\nu\over |\vec q|}$ and
$E_s=\sqrt{p_s^2+m^2}$. Here, $p_s$ is the spectator momentum 
and $M_d$ the mass of the deuteron.
The difference between $\tilde p_s^z$ and $p_z$ accounts for the 
longitudinal momentum transfer.
Spin indices  are suppressed to simplify the notations. The 
function ${\bf F^{np}}$ represents the FSI between the outgoing 
baryons and its form depends on the model describing the soft rescattering.
Within the elastic eikonal (Glauber) approximation (GA),
${\bf F^{np}(\vec k)}\rightarrow f^{np}({\bf \vec k_t})$, where 
$f^{pn} =\sigma^{pn}_{tot}(i+a_n)e^{-b_n k_t^2/2}$. 
At Q$^2>3$ (GeV/c)$^2$, the quantities $\sigma^{pn}_{tot}$, $a_n$ and 
$b_n$ depend only weakly on the momentum of the knocked-out
nucleon, with $\sigma^{pn}_{tot}\approx 40$ mb, $a_n\approx -0.2$ and
$b_n\approx~6-8$ GeV$^{-2}$ for the kinematics we use.

The reduced interaction  between 
the PLC and the spectator nucleon can be described 
in terms of its transverse size and the distance $z$ from the 
photon absorption point, i.e., in Eq.(\ref{eq:amp}) we replace  
${\bf F^{pn}}\rightarrow f^{PLC,N}(z,k_t,Q^2)$. For
numerical estimates of the reduced FSI $f^{PLC,N}(z,k_t,Q^2)$,
we use the quantum diffusion model (QDM) \cite{FLFS} as well as the three 
state model \cite{FGMS93}. Latter is based on the assumption 
that the hard scattering operator acts on a nucleon and produces a  
PLC, which is represented as a superposition of three baryonic states, 
$|PLC\rangle = \sum_{m=N,N^*,N^{**}} F_{m,N}(Q^2) |m\rangle$. 
In Fig.1, we compare the predictions of the elastic eikonal and the
two CT models for the transparency, $T=\sigma^{FSI}_{e,e'p}/
\sigma^{PWIA}_{e,e'p}$, for an unpolarized target.
We consider
so-called perpendicular kinematics, where the light cone 
momentum $\alpha={E_s-p^s_s\over m} \approx 1$ and $p_t\le
400~MeV/c$. It was demonstrated in Ref.\cite{FGMSS95} that these
kinematics maximize the contribution from the FSI and minimizes various
theoretical uncertainties.
One can see from Fig.1 that, optimistically,
one may expect $30\%$ effects from CT at $Q^2 \ge 4-6$ GeV$^2$.

Using a polarized target 
emphasizes the role of the deuteron's $D$-state,
allowing to probe the space-time evolution at 
smaller space-time intervals.
For numerical estimates, we consider the asymmetry $A_d$ measurable in 
electrodisintegration of a polarized deuteron with
helcities of $\pm 1$ and 0: 
$A_d(Q^2,\vec p_s) = {\sigma(1)+\sigma(-1)-2\cdot\sigma(0)
\over \sigma(1)+\sigma(0)+\sigma(-1)}$,
where $\sigma(s_z)\equiv{d\sigma^{\vec s,s_z}\over dE_{e'} d\Omega_{e'} 
d^3p}$ and $s_z$ is the deuteron's helicity. 
The $Q^2$ dependence of the asymmetry $A_d$  for "perpendicular"
kinematics, at $p_t=300$ MeV/c, is presented in Fig.2. 
One can see from this figure that CT effects 
can change $A_d$ by as much as factor of two for 
$Q^2\sim 10$ GeV$^2$.

\medskip

\noindent{\Large\bf 3 \ Study of the Relativistic Effects}

Let us consider now different kinematics, namely $Q^2\le 4$ GeV$^2$. 
In this case we expect minimal CT effects and therefore the consequences
of the FSI are well under control.  The kinematics, where 
the light-cone momentum $\alpha>1$ and $p_t\approx 0$,
are most sensitive to relativistic effects in the deuteron.
There are several techniques to treat the deeply bound nucleons as well as
relativistic effects in the deuteron. One group of approaches 
handles the virtuality of the bound nucleon within a
description of the deuteron in the lab. frame (we will call them virtual
nucleon (VN) approaches) 
by taking the residue over the energy of the spectator nucleon.
One has to deal with negative energy states which arise  for 
non-zero virtualities (see e.g. Ref.\cite{MST}).  Due to the binding, current 
conservation is not automatic and one has to introduce a prescription to 
implement e.m. gauge invariance (see e.g. Ref.\cite{deF}). 
Another approach is based on the observation that high energy processes
evolve along the light-cone.  Therefore, it is natural to describe the 
reaction within the light-cone non-covariant framework \cite{rep}. 
Negative energy states do not enter in this case, though one has to take into 
account so called instantaneous interactions.
For this purpose one employs e.m. gauge invariance to 
express the ``bad'' electromagnetic current component 
(containing instantaneous terms) through the ``good'' component 
$J^A_+ = -q_+/q_-J^A_-$ \cite{rep}.
In the approximation when non-nucleonic degrees of freedom in the
deuteron wave function can be neglected, one can unambiguously relate
the light-cone wave functions to those calculated in the lab. frame
by introducing the LC $pn$ relative three momentum
$k=\sqrt{{m^2+p_t^2\over \alpha(2-\alpha)} - m^2}$.

Turning to numerical estimates, it is worth noting that it is well 
established that, by using a polarized deuteron target in $(e,e'p)$
reactions, one can decisively disentangle the VN and LC 
prescriptions (see e.g. \cite{rep}). Now using the recent advances in
the FSI calculation, one can repeat a similar comparison for the
tensor asymmetry, $T^{20} = {1\over 3}(\sigma^{1,1} + \sigma^{1,-1} 
- 2 \sigma^{1,0})$, accounting also for the FSI diagrams. The result of such a 
comparison is presented in  Fig.3 for backward kinematics ($\theta_{s}=180^o$).
One can see that account of the FSI further increases the difference between 
the predictions of the VN and LC approaches, thus making their
experimental investigation more feasible.  

The advantage of using a $\vec d$ target to enhance
the contribution of small internucleon distances
\newpage

\begin{figure}[t]
\vspace{1.5cm}
\centering{
\epsfig{figure=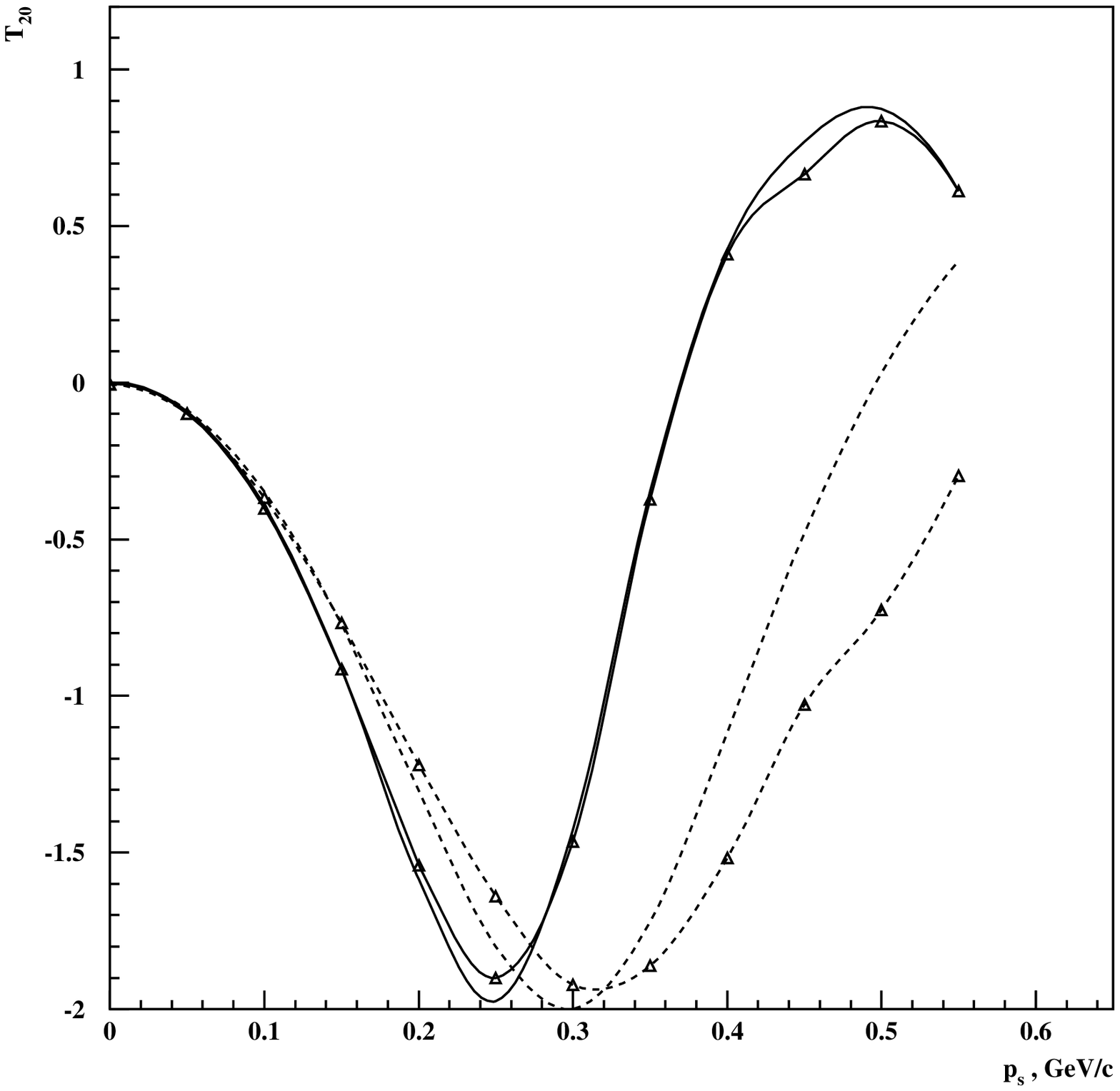,height=8.0cm}}
\end{figure}
\begin{figure}[t]
\vspace{-10.3cm}
\hspace{8.4cm}
\centering{
\epsfig{figure=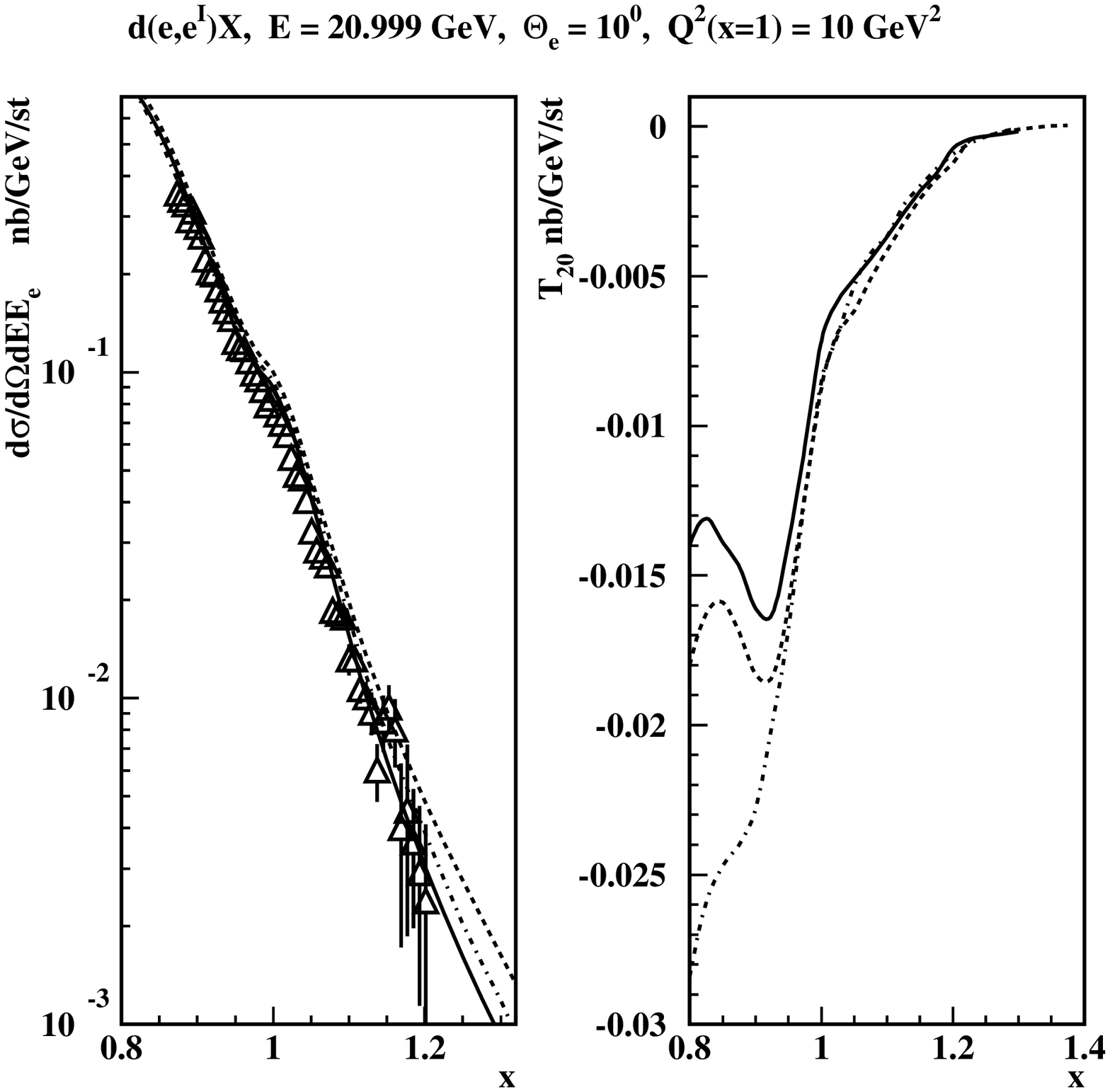,height=8.0cm}}
\end{figure}

\begin{picture}(100,200)(0,0)
\put(-15,210){{\bf Figure 3:} {\em $p_s$ dependence of the $(e,e'p)$}}
\put(-15,195){{\em tensor polarization at $\theta_{s}=180^0$. Solid}}    
\put(-15,180){{\em and dashed lines are PWIA predictions}}
\put(-15,165){{\em of the LC and VN methods, respective}}
\put(-15,150){{\em marked curves include FSI.}}

\put(235,210){{\bf Figure 4:} {\em $Q^2$ dependence of the unpolarized}}  
\put(235,195){{\em and tensor polarized cross sections. Solid }}
\put(235,180){{\em line - LC approach with PLC suppression,}}
\put(235,165){{\em dashed - LC, and dashed-dotted - VN.}}
\put(235,150){{\em Experimental data from Ref.\cite{Rock}.}}
\end{picture}

\vspace{-5.0cm}

\noindent 
holds even for inclusive $\vec d(e,e')$ scattering.
In Fig.4, we compare the predictions of the VN and CT
approaches for $d(e,e')$ reactions with unpolarized and polarized
deuteron targets. Yielding practically the same predictions for a unpolarized
target at $x<1$, the two approaches differ by as much as a factor of two 
in the tensor polarization cross section.

\medskip

\noindent{\Large\bf 3 Conclusions}

We demonstrated that the use of a polarized deuteron target allows to 
probe effectively smaller internucleon distances in the deuteron
ground state wave function for semiexclusive $(e,e'N)$ and inclusive 
$(e,e')$ reactions. This opportunity can be successfully used to 
gain a better understanding of the structure of (moderate) high energy,
large $Q^2$ $eA$ interactions.  In particular, we demonstrated
that the use of a $\vec d$ target would allow to observe 
the onset of Color Transparency at intermediate energies as well as to 
confront different descriptions of relativistic effects in the deuteron and 
electromagnetic interactions with deeply bound nucleons.

\end{document}